# Single Scan Dual-energy Cone-beam CT Using Static Detector Modulation: A Phantom Study

Junbo Peng

Nuclear & Radiological Engineering and Medical Physics Programs,
Woodruff School of Mechanical Engineering,
Georgia Institute of Technology, Atlanta, GA, 30332 USA

**Abstract**

Cone-beam CT (CBCT) is installed in the treatment room to facilitate online clinical applications, including image guidance in radiation and surgery. Half-fan and short-can are the commonly used modes in clinical applications to expand the imaging volume and reduce radiation exposure. Anatomical structures are faithfully reconstructed using CBCT while deep information including material composition is not feasible in its current form. Dual-energy computed tomography (DECT) is an advanced medical imaging technology with superior capability of material differentiation by scanning the patient using two different energy spectra. Conventional DECT systems require expensive and sophisticated hardware components, significantly limiting DECT applications to cone-beam CT (CBCT) machines. In this work, we propose a dual-energy cone-beam CT (DECBCT) system by placing a dedicatedly designed beam modulator in front of the detector surface to acquire dual-energy projection data in a single scan. A deep learning-based data restoration model is introduced to generate complete dual-energy data from the detector-modulated data for DECT image reconstruction and material decomposition. The performance of the proposed method is evaluated using an electron density phantom and clinical CT. The mean error for electron density calculation is lower than 1.16% in the electron density phantom study. The mean errors for reconstructed linear attenuation coefficient (LAC) are lower than 1.41% and 0.89% in the head and pelvis studies, respectively. The results indicate that the proposed method is a promising solution for single-scan DECBCT in clinical practice.

Keywords: Dual-Energy CBCT, Detector Modulation, Filtered Sinogram Restoration, Material Decomposition

## 1. Introduction

As a variant of diagnostic CT, cone-beam CT (CBCT) is installed in the treatment room to facilitate online clinical applications, including image guidance in radiation and surgery. To utilize the limited operating space in the treatment room, the CBCT system requires a systematic architecture that is low-cost, compact, open-gantry, and lightweight. Distinct from the imaging components in a diagnostic CT, flat-panel detector and low-power x-ray source are used as the major components in the CBCT system. To fully utilize the limited area of a flat panel detector without increasing hardware cost, a half-fan scan mode is designed to image objects with a greater field of view (FOV). This is achieved by shifting the detector along its horizontal plane to one side by less than half of its length for large-volume CBCT imaging. After the shift, the object is not fully detected within one projection. To compensate for the missing data, a full scan of 360 degrees is required to acquire sufficient data for accurate reconstruction. The Varian's CBCT pelvic mode, which is installed in its main-stream linear accelerator, is an example of this CBCT configuration. Another mode is referred to as short scan to image small-FOV object in less than 360 degrees which is the 180 degrees plus a fan angle spanned by the x-ray source. This scheme requires no offset of the detector position and can reduce the scan time and dose to patients. In both modes, the data redundancy is much less than that in the full-fan 360-degree scan which imposes higher difficulty in image reconstruction.

Dual-energy computed tomography (DECT) is a type of spectral CT that explores x-ray energy-dependent attenuation property of different materials, hence improving the capability of materials differentiation when comparing with conventional CT. Using dual-energy datasets, the material decomposition technique decomposes the CT images into basis material images including volume fraction maps which represent the compositions of imaged objects. In clinical practices, DECT and material decomposition are widely used for automatic separation of bone and contrast-agent-filled vessels in the angiographic scans, iodine quantification (Chandarana *et al.*, 2011), pseudo-monochromatic image generation (Yu *et al.*, 2011), and virtual non-enhanced imaging (Sommer *et al.*, 2012). To take advantage of these advanced functions in image guidance, dual-energy CBCT

(DECBCT) has been studied in prior work for material classification and improved dose calculation (Zbijewski et al., 2014; Men et al., 2017; Li et al., 2012).

Commercial DECT can be implemented using various schemes, including source-side and detector-side solutions. The source-side methods include double-scan, dual-source, and fast kVp-switching techniques (Zhang et al., 2010; Matsumoto et al., 2011). Double scanning at two different x-ray spectra acquires two sets of complete projection data. Nevertheless, this technique requires a long data acquisition time and has inconsistent geometry, which may lead to the geometric displacement between the two dual-energy scans. Physiological motions become a concern during the long acquisition time, as subsequent artifacts are introduced on decomposed images, especially for abdominal, chest, and cardiac imaging. Specialized scanners with sophisticated hardware components (e.g., dual sources (Zhang et al., 2010) and a fast kVp-switching x-ray source (Matsumoto et al., 2011)) can perform dual-energy imaging with scan time and dose exposure that are comparable to those of a standard CT scan.

The detector-side methods include dual-layer (Pelgrim et al., 2017) and photon-counting detector schemes (Wang et al., 2011). Both solutions acquire complete projection datasets at different energy levels in a single scan, but each has its limitations. The dual-layer detector absorbs low- and high-energy photons at the top and bottom layers of its scintillators, respectively. Nevertheless, the spectral separation of x-ray photons between the two layers is not significantly different, leading to a high condition number of the decomposition system and a highly unstable decomposition problem (Niu et al., 2014). The photon-counting detector of spectral CT records x-ray photons at different energy levels. This technology experiences challenges that are yet to be resolved, including electronics of large data acquisition, high counting rates, and large-area crystal growth (Thuering et al., 2019). In the aforementioned commercial DECT solutions, complete projection data are acquired at different spectra to facilitate the engineering implementation of dual-energy reconstruction and material decomposition.

The requirement for an ideal clinical implementation of DECBCT imaging includes minor modification to the existing CBCT hardware, no extra radiation dose to the patient, and an efficient reconstruction algorithm. As such, these commercial DECT imaging schemes are not readily available to be directly applied on DECBCT due to these systematic and online imaging requirements. A promising solution to meet all these strict constraints is the primary beam modulation method, which is an economical and effective way for dual-energy data acquisition. The primary modulation method places a series of metal strips at the exit of the x-ray tube to selectively harden x-ray beams in specific paths. This method is divided into two categories. The first type applies a strip-moving strategy to produce a uniform sampling density map which is advantageous for the compressed-sensing (CS) inspired algorithm (Lee et al., 2017; Cho et al., 2020). An additional mechanical device is installed to synchronize the modulator reciprocation with the x-ray exposure and data acquisition during the CBCT scan, though its clinical practicability may be degraded by the complicated mechanical control. An inconsistent data acquisition pattern is observed in the scans with and without objects in the field of view, leading to inaccurate calculation of line integrals. To combat the issues using a moving modulator, a stationary modulator is designed with shorter strip width than the spacing to acquire sufficient low-energy data to reconstruct high-quality reference low-energy CBCT images (Petrongolo and Zhu, 2018). A model-based CS reconstruction scheme using image similarity based regularization is applied to compensate for the high-energy data loss. Still, the stationary strategy causes severe high-energy data loss, which may degrade the accuracy of material decomposition.

Primary data loss is the major bottleneck to the accurate dual-energy reconstruction and material decomposition using the primary modulation methods. The sources of the data loss are mainly two folds. One source is the huge geometric magnification factor due to the short source-to-modulator distance. The transition area of the strip boundary is thus amplified by around six times on the detector plane, leading to significant primary data loss by up to 40%. The other source of data loss is the sparse sampling principle within a single CT scan to alleviate the primary loss. As such, the modulated dual-energy projection data are acquired using 360-deg full scan and reconstructed using iterative scheme with the CS regularization to suppress ring and streaking artifacts due to sparse sampling. Accurate reconstruction is more difficult using clinical scanning modes including the half-fan and short scans where only about 55% of full-scan data are acquired. To compensate for the primary data loss, various schemes have been developed in the sinogram restoration. They are divided into two categories, based on the techniques applied in these schemes: the interpolation-based and the learning-based schemes. Interpolation-based strategies include a linear (Brooks et al., 1978), an intensity directional (Bertram et al., 2009; Zhang and Sonke, 2013), and a partial differential equation schemes (Kostler et al., 2006). The performance of interpolation-based schemes heavily relies on the under-sampling pattern of the measured projection data. Residual artifacts are observed in the reconstructed images in the engineering practice using these methods. Deep learning-based methods are emerging strategies in recent years that are inspired by the rapid development of machine learning techniques. They have shown great superiority over the interpolation-based methods in the sinogram restoration tasks in various aspects, including the sparse-view (Lee et al., 2018), limited-angle (Wang et al., 2020), and slow kVp-switching (Cao et al., 2021) reconstruction scenarios.

In this work, we develop a novel modulation method, which meets the requirements of DECBCT imaging and avoids the shortcomings of existing schemes. The modulator is placed in front of detector instead of x-ray exit to suppress the primary data loss due to large penumbra areas from high magnification on strip boundaries. The strips of the modulator are placed in the axial direction to generate dual-energy data within each slice. As such, strip widths with equal spacings are built to acquire balanced dual-energy data. Higher modulation frequency can be applied to achieve fine structure sampling within DECBCT data. The strip width can also be enlarged by one order of magnitude than that in the primary modulator, and significantly reduces the complexity of mechanical manufacture. With the novel detector modulator, the projection data at two different energy spectra can be acquired in a single scan with suppressed data loss due to boundary penumbra effects. The missing spectral projection data in the modulated acquisition are restored by a neural network that learns the complete anatomical structure and balanced spectral information within the detector-modulated single-scan data. After the missing data restoration, full projections are estimated within each spectrum for efficient reconstruction and material decomposition using the fast iterative shrinkage-thresholding algorithm (FISTA). The performance of the proposed method is evaluated using an electron density phantom in an experimental study and clinical CT data in a simulation study.

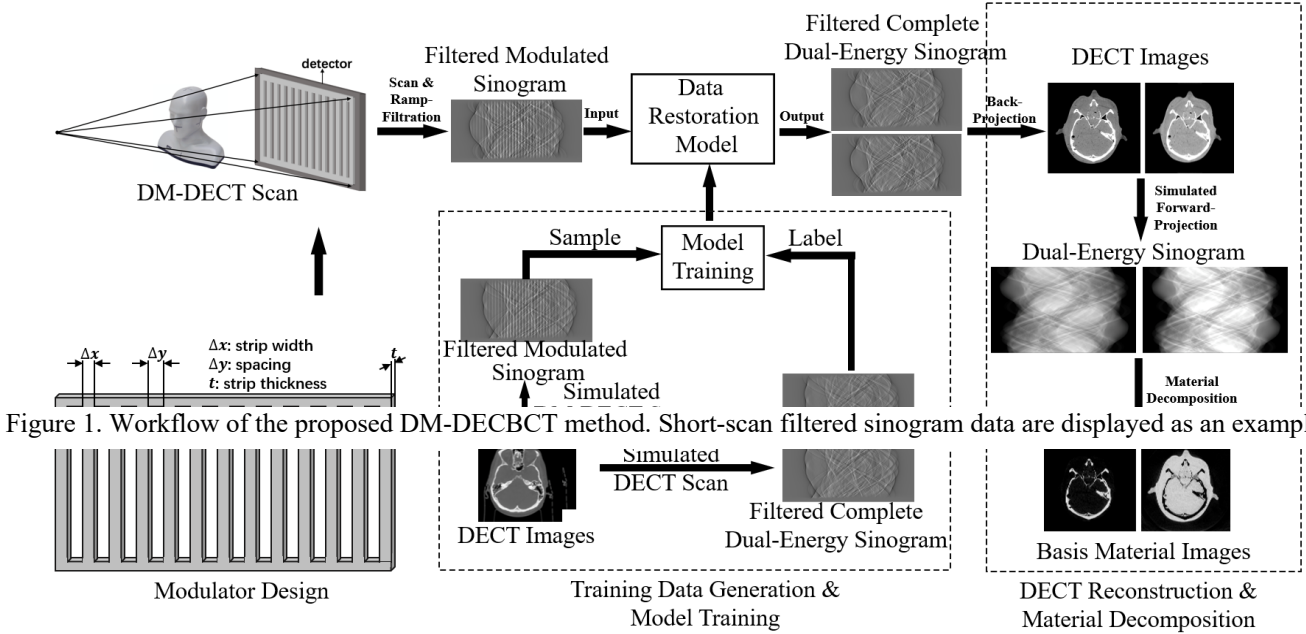

Figure 1. Workflow of the proposed DM-DECBCT method. Short-scan filtered sinogram data are displayed as an example.

## 2. Methods and Materials

### 2.1 Workflow

The workflow of the proposed DECBCT imaging method is shown in Fig.1. A modulator composed of a series of metal strips is installed in front of the flat-panel detector to selectively harden the x-ray beams that are passing through those strips. These metal strips are distributed along the axial direction of the detector to generate dual-energy projections regardless of the cone-angle value. The total area of the projected strips is half of that on the detector to achieve balanced data fidelity for dual-energy data acquisition. The material composition and thickness of the strips are selected according to the electron density calculation accuracy of an evaluation phantom to ensure sufficient energy spectra separation. The proposed spectrum modulator generates two groups of x-ray beams with distinct energy spectra, enabling data acquisition of under-sampled dual-energy projection at each view within one single scan.

Due to the under-sampled spectral data acquisition in the half-fan and short-scan modes, the detector-modulated projection data is insufficient for both DECT reconstruction and material decomposition. When analyzing the acquired data, the anatomical structure is observed to be fully preserved in the spectra-mixed data, while the LAC information is only partially acquired in both spectra for its energy-dependent property. The data restoration operation is then introduced to generate complete dual-energy data through a pix2pix GAN model, which extracts the complementary anatomical information from one of the spectral data sets to compensate for the missing information of its counterpart.

Considering the large quantity required for training data, the two-scan and detector-modulated DECT (DM-DECT) data are simulated as the label and sample data, respectively. Instead of direct restoration in the projection or sinogram domain, a high-pass ramp-filtered sinogram is utilized because of its derivative property (Noo *et al.*, 2004; Zeng, 2007) and reduced artifacts on reconstructed images. With the restored complete dual-energy data, DECT images can be directly reconstructed using standard back-projection operation. Then the ring correction method (Sijbers and Postnov, 2004) is employed to eliminate the residual artifacts on the reconstructed images. As an inverted ramp filtration is not available, complete dual-energy sinograms can be obtained by forward-projection operation on the reconstructed DECT images. The proposed multi-material decomposition is implemented in the sinogram domain due to its relatively smooth spatial distribution after line integral operation compared with CT image.

## 2.2 Strip material and thickness selection

The spectral separation and signal-to-noise ratio (SNR) of the acquired data are determined by the material composition and strip thickness of the modulator. A larger spectra separation is preferred for stable material decomposition at the expense of an increased noise level and decreased accuracy of quantitative imaging.

Table 1. Spectral shift (keV) and corresponding metal thickness (mm) at different attenuation ratios by varying filtration materials, assuming 10 cm liquid water in the beam. Boone's method (Boone and Seibert, 1997) is used to generate the spectra.

| Spectral Shift(keV)/ thickness (mm) Materials | 20% | 40% | 60% | 80% |
|---|---|---|---|---|
| Al | 1.45/3.00 | 3.17/7.02 | 5.40/12.95 | 8.83/23.70 |
| Pb | 0.50/0.205 | 0.99/0.048 | 1.49/0.0905 | 1.98/0.17 |
| Cu | 2.64/0.16 | 6.00/0.415 | 10.12/0.81 | 16.57/1.65 |
| Mo | 2.76/0.054 | 6.13/0.1355 | 10.63/0.273 | 17.67/0.572 |
| Sn | 2.74/0.05 | 6.13/0.125 | 10.63/0.25 | 17.69/0.52 |

Material composition is optimized according to the spectral shift of x-ray beams before and after passing through different materials when attenuating the same quantity of photons. Specifically, an optimal filtration material should have a higher energy shift when the transmission ratio is fixed. In this work, the spectral separation is quantified using the mean energy shift of x-ray beams and the results are summarized in Table 1. The unmodulated x-ray beam is generated using 125 kVp tube voltage with an inherent 0.7 mm Al and an additional 0.2 mm Cu filters to reduce beam-hardening artifacts. Spectrum simulation is performed using Boone's method . The five commonly used modulator materials are aluminum, lead, copper, molybdenum, and tin. The three latter on this list are preferred since they outperform the two former materials with more significant spectral separations. Nevertheless, the copper filter is not as ideal as molybdenum and tin as it is three times thicker than the other two. No obvious difference of spectral shift is observed between molybdenum and tin filters. As such, both molybdenum and tin are optimal candidates for spectral modulators.

For a specific modulation material, a thicker strip results in a larger spectral separation between high- and low-energy data and lower SNR on the high-energy projection data. The former effect is desired to achieve a larger condition number of material composition matrix, while the latter effect is detrimental for precise quantitative imaging. The optimal strip thickness should achieve a balanced tradeoff between the increased spectral separation and decreased SNR of acquired data, which will yield accurate material decomposition. In this work, the optimal thickness of modulator thickness is determined using material decomposition and electron density calculation performed on the electron density phantom with the model number of 062M from Computerized Imaging Reference Systems (CIRS) .The material composition and thickness leading to the most accurate electron density calculation is selected to build the detector modulator.

## 2.3 Filtered Complete Sinogram Restoration

### 2.3.1 Network architecture and model training

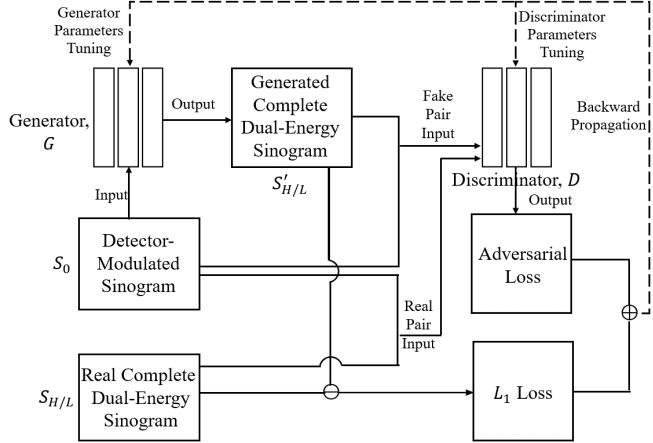

Figure 2. The architecture of the model training for filtered sinogram restoration.

Due to its success on similar sinogram restoration tasks (Cao *et al.*, 2021), the pix2pix GAN model, a supervised approach to implement image-to-image translation tasks, is applied to build the data restoration model for filtered sinogram, and to learn the mapping from the spectra-mixed data using detector modulation to complete dual-energy data. The network employs a complex loss function including a conventional conditional GAN loss and a $L_1$ loss to preserve both the low- and high-frequency information within the image restoration, respectively (Isola *et al.*, 2017). In this work, high- and low-energy data are trained separately using the same network structure with individualized training sets.

The architecture of the model training is shown in Fig. 2. $S_0$ is the detector-modulated sinogram after ramp filtration as sample data. $S_{H/L}$ represent the filtered complete high- or low-energy sinogram acquired using simulated DECT scan as the label. The generator $G$ produces filtered complete high- or low-energy sinogram, $S'_{H/L}$, from the detector-modulated data, $S_0$. $S'_{H/L}$ and $S_0$ compose the synthetic pair for the discriminator. The real filtered sinogram, $S_{H/L}$, and the detector-modulated data, $S_0$, are combined into the discriminator as the real pair. The discriminator acts as a classifier to distinguish the real image pair from the synthetic ones and generates an adversarial loss, $\mathcal{L}_{cGAN}$. The $L_1$ distance between the synthetic filtered sinogram, $S'_{H/L}$, and real data, $S_{H/L}$, is calculated to generate the residual part of the loss function, $\mathcal{L}_{L1}$. Hence, the total loss function for parameter optimization is written as:

$$\mathcal{L}_{total} = \mathcal{L}_{cGAN} + \lambda \cdot \mathcal{L}_{L1}, \quad (1)$$

where $\lambda$ is a tuning factor to balance the tradeoff between the restoration of high- and low-frequency structures (Isola *et al.*, 2017). The dashed lines represent the feedback from the total pix2pix GAN loss to the generator and discriminator for parameter optimization.

2.3.2 Generator and discriminator

The structure of the generator, $G$, as displayed in Fig. 3 (a), transforms the spectra-mixed filtered sinogram, $S_0$, to complete filtered sinogram data, $S'_{H/L}$. The ResNet-based encoder-decoder architecture, which is reported to produce better results at reduced computation cost (Li *et al.*, 2020), is applied to build the generator in this work. Three convolutional layers are employed in the encoder. The quantity of residual blocks (Fig. 3c) is set as six to balance the tradeoff between improved performance and increasing computational cost. Convolutional layers extract structural and textural information of the input images. The residual blocks prevent gradient vanishing while allowing the learning process of identity mapping easier (Johnson *et al.*, 2016). The decoder part consists of three symmetrically transposed convolutional layers with the encoder part to generate an image with the same size as the input. The activation function in the final layer is tanh function with an output range of [-1 1], which can be easily converted into any pixel values using an affine transformation (Karlik and Olgac, 2011).

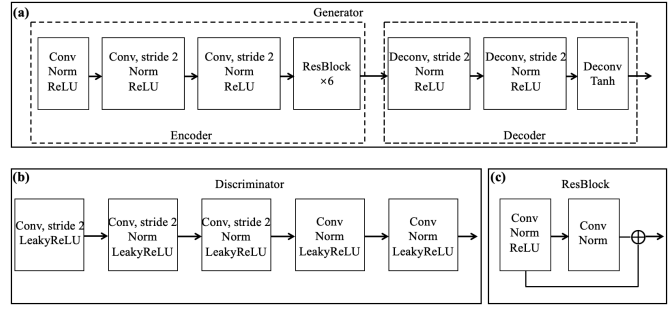

Figure 3. The structure of each module in the pix2pix GAN model.

The discriminator, $D$, is constructed using a commonly used PatchGAN to distinguish the fake from the real pair. It takes both the source image and the target image as input and predicts the likelihood of whether target image is a real or fake translation of the source image. As shown in Fig. 3(b), the PatchGAN discriminator is composed of five convolutional layers to map the input to a probability matrix. Each element in the matrix represents the adversarial loss of a patch in the image. All the loss values are averaged to produce the ultimate output of the discriminator (Isola *et al.*, 2017). The patch-based discriminator is introduced to model the high-frequency structures and overcome the image blurring effect of the additional $L_1$ loss (Larsson *et al.*, 2016). With the generated filtered complete sinogram, $S'_{H/L}$, DECT images can be readily reconstructed using direct back-projection operation. The complete dual-energy sinograms can be obtained by forward projecting the DECT images.

2.4 Material Decomposition

The DECT images are reconstructed using model-generated complete data. The complete dual-energy sinograms ($\vec{b}_H$ and $\vec{b}_L$) can be obtained using the forward projection operation according to corresponding half-fan or short-scan geometry. An image-domain three-material decomposition is implemented in an iterative framework directly from the dual-energy sinogram data. The optimization problem can be formulated as

$$\min_{\vec{x}} \phi(\vec{x}) = \frac{1}{2}\left\|FA\vec{x} - \vec{b}\right\|_2^2, \quad (2)$$

$$s.t. \ \vec{x} \succcurlyeq 0, \sum_i \vec{x}_i = \vec{1}, \quad (3)$$

$$F = \begin{bmatrix} FP & O \\ O & FP \end{bmatrix}, A = \begin{bmatrix} \mu_{1,H} & \mu_{2,H} & \mu_{3,H} \\ \mu_{1,L} & \mu_{2,L} & \mu_{3,L} \end{bmatrix},$$

$$\vec{x} = \begin{bmatrix} \vec{x}_1 \\ \vec{x}_2 \\ \vec{x}_3 \end{bmatrix}, \vec{b} = \begin{bmatrix} \vec{b}_H \\ \vec{b}_L \end{bmatrix},$$

where $FP$ indicates the forward-projection operator, $A$ is the composition matrix consisting of linear attenuation coefficients (LACs) of different basis materials at two spectra ($\mu_{1/2/3,H/L}$). $\vec{x}$ is the volume fraction of basis images, which is the vectorized maps of all basis materials. $\vec{x}$ meets the sum to one and the box constraints, according to the volume or mass conservation property. This optimization problem can be converted into an equivalent form by replacing the constraints with an indicator function added to the objective function (Yang *et al.*, 2017) as shown in the following

$$\min_{\vec{x}} \phi(\vec{x}) + \Gamma(\vec{x}), \quad (4)$$

$$\Gamma(\vec{x}) = \begin{cases} 0, & \text{if } \vec{x} \succcurlyeq 0, \Sigma_i \vec{x}_i = \vec{1} \\ \infty, & \text{otherwise} \end{cases}. \quad (5)$$

The unconstrained optimization problem can be solved using the FISTA algorithm with line search due to its high computational efficiency (Beck and Teboulle, 2009; Yang *et al.*, 2017). In all cases, the third basis material is set as air in the following cases to achieve the volume conservation. The pseudo code is shown in Table 2.

Table 2. Multi-material decomposition from dual-energy sinograms using FISTA algorithm with line search

1 **Initialization:**

2 $\vec{x}_0 = \vec{0}, \vec{v}_0 = \vec{x}_0, r_1 = 0.8, r_2 = 0.2, t_0 = 1, N_{max} = 300$

3 **For** $n = 1: N_{max}$ **do**

4 $\quad t := t_{n-1}/r_1$

| 5 | **Repeat** |
| --- | --- |
| 6 | $\theta := \begin{cases} 1 & \text{if } n = 1 \\ \text{positive root of } t_{n-1}\theta^2 = t\theta_{n-1}^2(1-\theta) & \text{if } n > 1 \end{cases}$ |
| 7 | $\vec{y} := (1-\theta)\vec{x}_{n-1} + \theta\vec{v}_{n-1}$ |
| 8 | $\nabla\phi(\vec{y}) = A^T F^T (FA\vec{y} - \vec{b})$ |
| 9 | $F(\vec{y}) = \phi(\vec{x}) + \Gamma(\vec{y})$ |
| 10 | $\vec{y}' = \vec{y} - t \cdot \nabla\phi(y)$ |
| 11 | $\Delta\vec{y}' = (\Sigma_i \vec{y}'_i - \vec{1})/3$ |
| 12 | $\vec{x} := \max(\vec{0}, \vec{y}' - \Delta\vec{y}')$ |
| 13 | $F(\vec{x}) = \phi(\vec{x}) + \Gamma(\vec{x})$ |
| 14 | **break if** $F(\vec{x}) \le F(\vec{y}) + \langle \nabla F(y), \vec{x} - \vec{y} \rangle + \frac{1}{2}\|\vec{x} - \vec{y}\|_2^2$ |
| 15 | $t := r_2 t$ |
| 16 | $t_n := t$ |
| 17 | $\theta_n := \theta$ |
| 18 | $\vec{x}_n := \vec{x}$ |
| 19 | **break if** $\|\vec{x}_n - \vec{x}_{n-1}\|_2^2 \le 1 \times 10^{-4}$ |
| 20 | $\vec{v}_n := \vec{x}_{n-1} + \frac{1}{\theta_n}(\vec{x}_n - \vec{x}_{n-1})$ |
| 21 | **End** |

## 2.5 Evaluation

### 2.5.1 Data acquisition

The proposed DM-DECT method is evaluated using both physical phantom and clinical CT simulation data. In the physical phantom study, an electron density phantom is used to verify the efficacy in quantitative imaging task of the proposed method using the half-fan and short scan modes. In the clinical study, a head and a pelvis CT data are utilized to demonstrate its feasibility in clinical imaging tasks using the short scan and half-fan protocols, respectively.

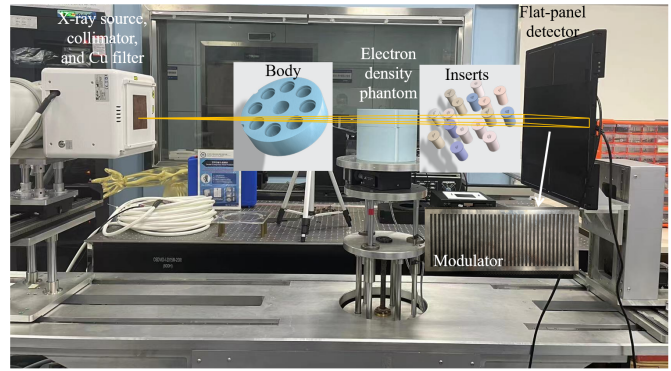

Figure 4. The proposed DM-DECT scheme on a table-top CBCT system. The manufactured detector modulator and the electron density phantom are shown in the top and bottom of the figure.

The electron density phantom is scanned on a customized table-top CBCT system with a source-to-detector distance of 1500 mm and a source-to-axis distance of 1000 mm. The x-ray tube has an inherent filtration of 0.7 mm aluminum and a target angle of 14 degrees. Additionally, a 0.2 mm copper sheet is attached to the source exit to pre-harden the primary x-ray beams and reduce the beam hardening effect during CT scan. To eliminate scatter contamination on the projection data, the phantom is scanned with a narrowly opened collimator of 2.2 cm on the detector, which is equivalent to a fan-beam geometry. Phantom data are acquired in continuous exposure mode with a tube voltage of 125 kVp, a tube current of 5 mA, and a small focal spot of 0.4 mm. The flat-panel detector in the table-top CBCT system is composed of 1408-by-1408 pixels with a pixel dimension of 0.308-by-0.308 mm². The phantom is scanned using a full-fan geometry in a 360-deg rotation. The projection data is cropped along the view angles and detector pixels in the short-scan and half-fan study, respectively. The data acquisition parameters are summarized in Table 3.

Table 3. Parameters of data acquisition.

| | |
|---|---|
| SID / SAD | 1500 / 1000 mm |
| Tube Voltage / Current | 125 kVp / 5 mA |
| Inherent Filter | 0.7-mm Al |
| Additional Filter | 2-mm Cu |
| Geometry | Fan Beam |
| Detector Pixel Size | 0.308 mm |
| View Angle Increment | 0.6° |
| Reconstruction Size | 512×512 |
| Half-Fan Geometry | |
| Studies | Physical phantom and pelvis |
| View Angle Range | 0°~360° |
| Number of Projections | 600 |
| Short Scan Geometry | |
| Studies | Physical phantom and head |
| View Angle Range | 0°~204° |
| Number of Projections | 340 |

The electron density phantom is composed of a cylindrical body with nine holes and pluggable inserts inside the holes. Due to the varying combinations of inserts in this work, detailed information of the inserts are listed in the captions of the figures. To expand the diversity of training and validation data, the position and selection of inserts are combined to construct eight different electron density phantoms. All phantoms are scanned with and without the metal sheet wholly covering the detector surface to obtain complete high- and low-energy projection data to generate training sets of phantom study. Then reconstructed eight DECT image pairs are augmented to 800 image pairs by image rotation and translation. The label data are obtained by forward projecting the augmented 800 DECT image pairs and ramp filtration. The sample data are generated by concatenating interlaced high- and low-energy sinograms after ramp filtration. Among all the 800 image pairs, 700 of them are used for model training and the other 100 pairs constitute the validation dataset. To show the robustness of the proposed scheme, the same electron density phantom with different inserts combination is scanned using the manufactured detector modulator to generate testing data.

The clinical CT data is simulated using the identical geometry and spectral settings as those in the physical phantom study. Since the spectral distribution of commercial DECT equipment is inconsistent with that in our design, clinical CT images are segmented into different tissues including bone, muscle, and adipose according to their HU value ranges. The simulated DECT images are created by assigning the LAC values of the segmented three tissues that are measured in the physical phantom study. The spectral settings are thus kept consistent between the physical experiment and clinical simulation study.

2.5.2 Implementation details

In this work, DECT images and volume fraction maps for each basis material are reconstructed and decomposed with an image dimension of 512-by-512 pixels and pixel resolutions of 1-by-1 mm$^2$ for the pelvis study and 0.5-by-0.5 mm$^2$ for other studies. In the electron density phantom study, two inserts labeled as solid dense bone and liver are selected as basis materials. In the clinical CT study, bone and soft tissue are selected as basis materials. In all cases, the third basis material is fixed to air. Poisson noise is simulated in the clinical data studies. The photon counting of low-energy air scan is set as $1 \times 10^5$ in each detector pixel to approach the noise level measured in the physical phantom data. The line-integral values inside transition areas for simulation data are obtained using linear interpolation method.

Strips with equal width and spacing are employed in this work to realize uniform modulation frequency across the whole object. In this work, 5-mm strip width is empirically determined to construct the beam modulator, which has a higher modulation frequency and reduced penumbra areas compared to previous beam-modulation works (Petrongolo and Zhu, 2018; Lee *et al.*, 2017; Cho *et al.*, 2020).

The proposed model training is implemented on the PyTorch framework using the open-source official pix2pix project (Isola *et al.*, 2017). All the training and testing processes are implemented on a GPU workstation with a single GeForce RTX 2080Ti Turbo graphics card of 12 GB graphic memory. The training process consist of 150 epochs at a constant learning rate of $2 \times 10^{-4}$ and 250 epochs with linearly declined learning rate to zero. Batch size is set as one in all cases. The weighting of $L_1$ loss is tuned to 200 for the physical phantom study and 400 in the clinical data studies for the optimal performance of data restoration.

2.5.3 Image quality metrics

To verify the feasibility of the DM-DECT method, we compare the results of the proposed dual-energy-scan method with the results of conventional two-scan scheme, which is considered as the ground truth. To evaluate the performance of dual-energy CT imaging, one-dimensional (1D) plots of DM-DECT images and ground truth are compared. Besides, average percentage error of LAC values and the structural similarity (SSIM) are calculated.

Mean LAC values are measured inside a uniform ROI on the DM-DECT and ground truth images and the average percentage error is determined by

$$\mu(\%) = \left|\frac{\bar{\mu}_{ROI} - \bar{\mu}_{ROI}^{ref}}{\bar{\mu}_{ROI}^{ref}}\right| \times 100\%, \qquad (6)$$

where $\bar{\mu}_{ROI}$ and $\bar{\mu}_{ROI}^{ref}$ indicate the mean LAC values inside a uniform ROI on the DM-DECT and ground truth images.

The SSIM is a good approximation of perceptual image quality. It indicates the similarity of two images regarding structure, contrast, and luminance and is defined as follow.

$$SSIM = \frac{(2\bar{x}\bar{y} + C_1)(2\sigma_{xy} + C_2)}{(\bar{x}^2 + \bar{y}^2 + C_1)(\sigma_x^2 + \sigma_y^2 + C_2)}, \qquad (7)$$

where $\bar{x}$ and $\bar{y}$ indicate the mean values of the CT images generated by DM-DECT method and two-scan strategy, respectively; $C_1 = (0.01 \cdot L)^2$ and $C_2 = (0.03 \cdot L)^2$ are the small constants to avoid divergence of the division operation; $L$ is the dynamic range of the input images; $\sigma_x$ and $\sigma_y$ are the standard deviations of two CT images; $\sigma_{xy}$ is the covariance between $\sigma_x$ and $\sigma_y$.

To evaluate the quantitative imaging performance of the DM-DECT method, electron density quantification is performed in the physical phantom study. The electron density of each ROI is calculated as

$$\rho_e = \rho_{e,1} \cdot \bar{x}_1 + \rho_{e,2} \cdot \bar{x}_2 + \rho_{e,3} \cdot \bar{x}_3, \qquad (8)$$

where $\rho_{e,i}$ is the referenced electron density of the material $i$, and $\bar{x}_i$ is the mean volume fraction of material $i$ inside the ROI. For each ROI, the average percent error is determined using

$$E(\%) = \left|\frac{\rho_e - \rho_e^{ref}}{\rho_e^{ref}}\right| \times 100\%, \qquad (9)$$

where $\rho_e^{ref}$ is the reference value of electron density for each ROI provided in the phantom's user manual.

## 3. Results

3.1 Modulator Design

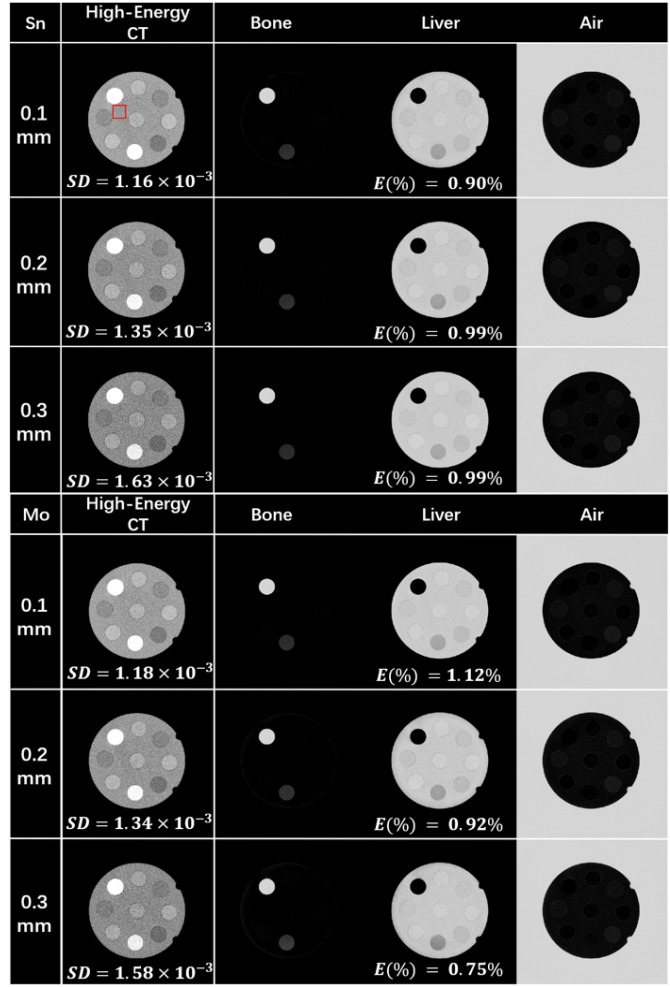

Figure 5. Reconstructed high-energy CT images and decomposed basis images using filters (Sn and Mo) with different thicknesses from 0.1 to 0.3 mm. Display windows are [0.015 0.025] mm-1 for CT images and [0 1.2] for basis images. SD is the standard deviation measured on a uniform ROI inside the dashed rectangular box. $E(\%)$ is the average percent error of electron density.

Based on the modulator selection in previous works (Lee *et al.*, 2017; Petrongolo and Zhu, 2018; Cho *et al.*, 2020), six metal sheets including tin and molybdenum from 0.1 to 0.3 mm act as candidates in the strip thickness optimization study. To perform material decomposition and subsequent electron density calculation, the electron density phantom is scanned twice - with and without the metal sheet covering the detector surface. For each set of complete dual-energy projection data, the proposed multi-material decomposition method is performed to generate volume fraction maps of the three basis materials. The average percentage error of electron density is calculated using the decomposed basis images. Reconstructed high-energy CT images are displayed in the first column in Fig. 5, showing varying noise levels with metal filters of different materials and thickness. The decomposed volume fraction maps of basis materials for the six metal sheets are displayed in the second to fourth columns in Fig. 5, where the error of electron density calculation for each candidate is indicated on associated basis images. No obvious artifacts or crosstalk are observed in all the decomposed basis images, and all the average percentage errors are close to 1%. Therefore, all the six beam modulator candidates are feasible for the quantitative imaging task. Taking the cost and manufacturing difficulty into consideration, 0.2 mm tin is selected as the optimal strip material and thickness in this work.

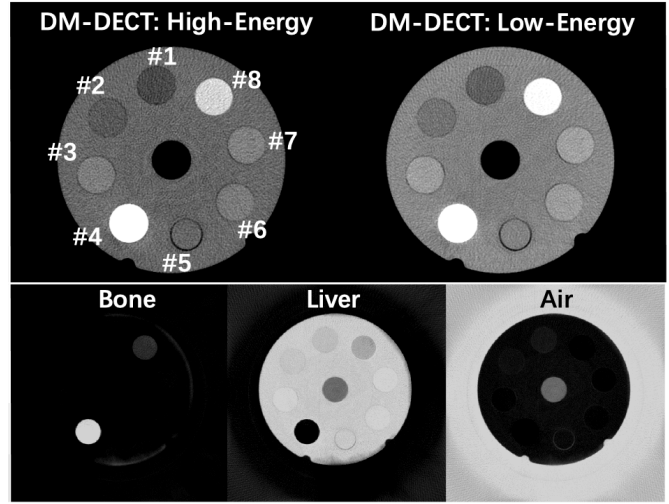

Figure 6. The first row: reconstructed DECT images using 5mm-strip modulator. The second row: decomposed basis images using the 5mm-strip modulator. Display windows are [0.015 0.025] mm$^{-1}$ for the CT images and [0 1.2] for the decomposed volume fraction images. The ROIs from #1 to #8 indicated on the high-energy CT image are adipose, breast, muscle, solid dense bone, removable vial, muscle, liver, and solid trabecular bone, respectively.

Before being put into production, a tin modulator with 0.2-mm thickness and 5-mm strip width is simulated to perform the DM-DECT imaging on the electron density phantom to verify its feasibility in practical use. The 5-mm strip width is selected based on the tradeoff between penumbra loss and spatial sampling of anatomical structure. Reconstructed DM-DECT images and decomposed basis images are displayed in Fig. 6, where no obvious artifacts appear on the DECT and basis material images. The electron density calculation results are listed on the Table 3 where the average error is lower than 1.2%, verifying its feasibility for the quantitative imaging task.

Table 3. Electron density (ED) calculation and corresponding percentage error of electron density calculation using different strip widths. The unit of electron density is $10^{23}$ cm$^{-3}$ and the error is in the form of (Mean±SD) %.

| ROI | #1 | #2 | #3 | #4 | #5 | #6 | #7 | #8 | Average Error |
|---|---|---|---|---|---|---|---|---|---|
| Ground Truth | 3.171 | 3.261 | 3.483 | 4.862 | 3.340 | 3.483 | 3.516 | 3.730 | |
| 5mm | 3.155 | 3.278 | 3.536 | 4.846 | 3.376 | 3.522 | 3.562 | 3.833 | |
| Error(%) | 0.51 ±2.45 | 0.53 ±2.37 | 1.52 ±1.24 | 0.34 ±0.50 | 1.06 ±2.35 | 1.12 ±1.55 | 1.29 ±0.71 | 2.76 ±0.87 | 1.14 ±1.51 |

3.2 Electron Density Phantom Study

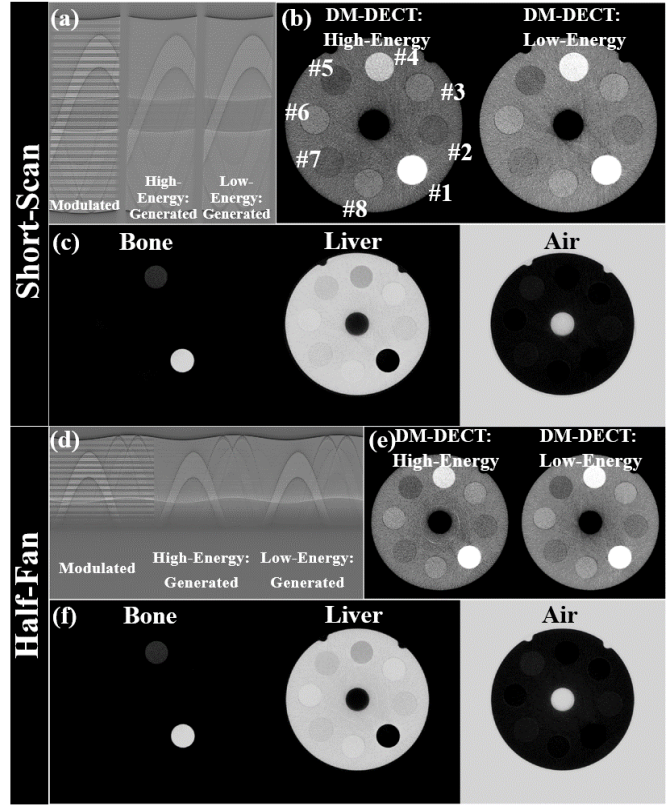

Figure 7. Short-scan and half-fan results in the physical phantom study: (a, d) the acquired modulated filtered sinogram, generated high- and low-energy filtered sinogram; (b, e) DECT images using the proposed DM-DECT method. (c, f) decomposed basis images of bone, liver and air using the proposed method. The display windows are adaptive for the filtered sinogram images, [0.015 0.025] mm$^{-1}$ for the DECT images, [0 1.2] for the decomposed volume fraction maps. The ROIs from #1 to #8 indicated in the high-energy CT image are solid dense bone, breast, liver, solid trabecular bone, adipose, liver, breast, and muscle, respectively. The inhomogeneous air zone in the decomposed are outside the FOV.

Installing a tin modulator with 0.2-mm thickness and 5-mm strip width on the detector surface, the detector-modulated dual-energy scan is performed on the table-top CBCT system using an electron density phantom. The phantom is scanned using full-fan geometry with 360-deg views. The acquired projection data are cropped to fit the half-fan and short-scan geometry. Fig. 7 shows the filtered detector-modulated, generated high- and low-energy sinograms, reconstructed DM-DECT images and decomposed volume fraction maps for each geometry. All the eight ROIs are preserved with complete structures and sharp boundaries on the DM-DECT images, showing the great performance of the proposed method for clinical tasks. No visible artifacts are observed on the decomposed images. The electron density calculation results are listed in Table IV. The mean percentage errors of short-scan and half-fan results are 1.12% and 1.16%, indicating the validation of the DM-DECT method for both the small- and large-FOV quantitative imaging scenarios.

Table 4. Half-fan results: Electron density calculation and its absolute percentage error in the physical phantom study. The unit of electron density is $10^{23}$ cm$^{-3}$ and the error is in the form of (Mean±SD) %.

| ROI | #1 | #2 | #3 | #4 | #5 | #6 | #7 | #8 | Average Error |
|---|---|---|---|---|---|---|---|---|---|
| Ground Truth | 4.862 | 3.261 | 3.516 | 3.730 | 3.171 | 3.516 | 3.261 | 3.483 | |
| Short-Scan | 4.854 | 3.285 | 3.476 | 3.782 | 3.162 | 3.502 | 3.277 | 3.508 | |
| Error(%) | 0.68 ± 1.76 | 1.19 ± 3.47 | 0.54 ± 1.92 | 1.84 ± 3.87 | 0.98 ± 3.26 | 0.79 ± 2.43 | 1.67 ± 3.84 | 1.28 ± 3.31 | 1.12 ± 2.98 |
| Half-Fan | 4.859 | 3.229 | 3.511 | 3.786 | 3.103 | 3.506 | 3.237 | 3.470 | |
| Error(%) | 0.83 ± 1.28 | 1.20 ± 3.61 | 0.72 ± 2.76 | 1.75 ± 3.95 | 1.77 ± 3.46 | 0.94 ± 1.44 | 1.02 ± 3.47 | 1.08 ± 2.10 | 1.16 ± 2.76 |

3.3 Simulated Short-Scan Head Study

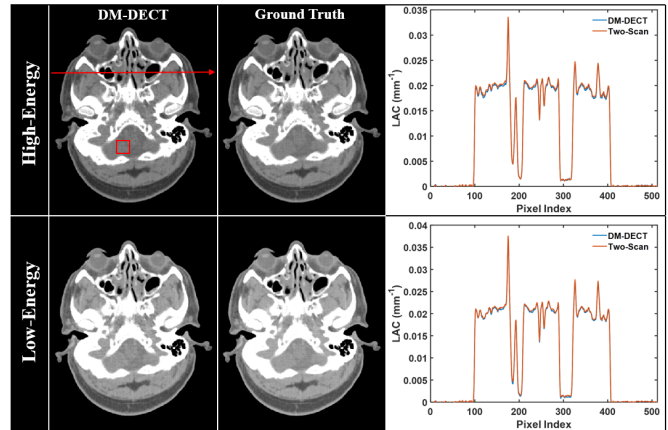

Figure 8. The first and second columns: DECT images using DM-DECT and two-scan method as the ground truth. The third column: line profiles along the red arrow within the high-energy DM-DECT image. Display windows are [0.015 0.025] mm-1 for all.

High- and low-energy CT images reconstructed using the DM-DECT method and the ground truth are displayed in the first and second columns of Fig. 8. The DM-DECT results exhibit high spatial resolution without obvious differences compared with the ground truth. The red box on the high-energy DM-DECT image indicates a uniform ROI for the average percentage error calculation. The errors are 1.05% and 1.41% for the high- and low-energy CT images. The SSIM values are calculated between DM-DECT images and ground truth, which both equal to 0.9999, showing the excellent structure-preservation ability of the proposed method.

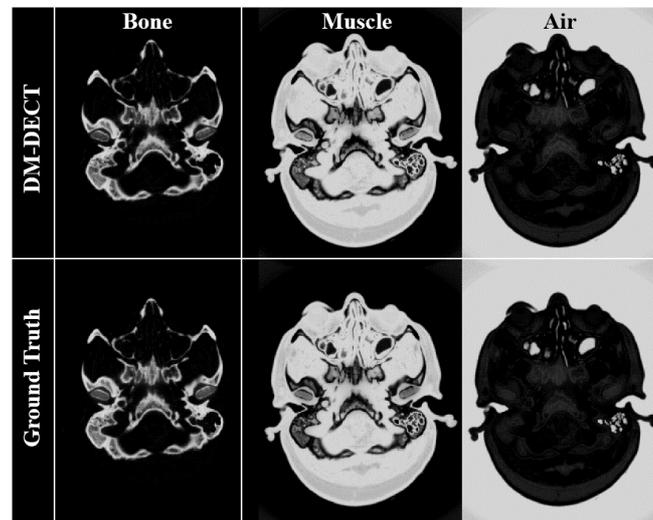

Figure 9. Decomposed volume fraction maps using DM-DECT and the two-scan method as the ground truth. Display windows are [0 1.2] for all.

Line profiles along the red arrows in the high-energy DM-DECT image are plotted on the third column of Fig. 8, showing great coherence in the reconstructed LAC values with the ground truth. Decomposed basis images using the DM-DECT method and the ground truth are displayed in Fig. 9, where no visible differences appear between the DM-DECT results and the ground truth. SSIM between the two sets of decomposed images is 0.9630, showing the great structure preservation performance of the DM-DECT method.

3.4 Simulated Half-Fan Pelvis Study

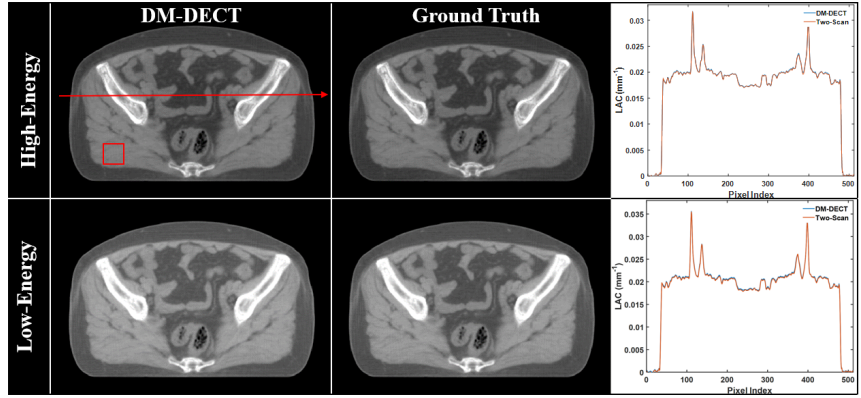

Figure 10. The first and second columns: DECT images using DM-DECT and two-scan method as the ground truth. The third column: line profiles along the red arrow indicated in the high-energy DM-DECT image. Display windows are [0.015 0.025] mm-1 for all.

Reconstructed DM-DECT images and the ground truth images are displayed in the first and second columns in Fig. 10, where the DM-DECT images show clear structure without visible differences compared with the ground truth. The red box indicates a uniform ROI for average percentage error calculation. The results are 0.84% and 0.89% for high- and low-energy CT images. The SSIM values are 1 and 0.9998 between DM-DECT images and the ground truth for the high-energy and low-energy results, respectively.

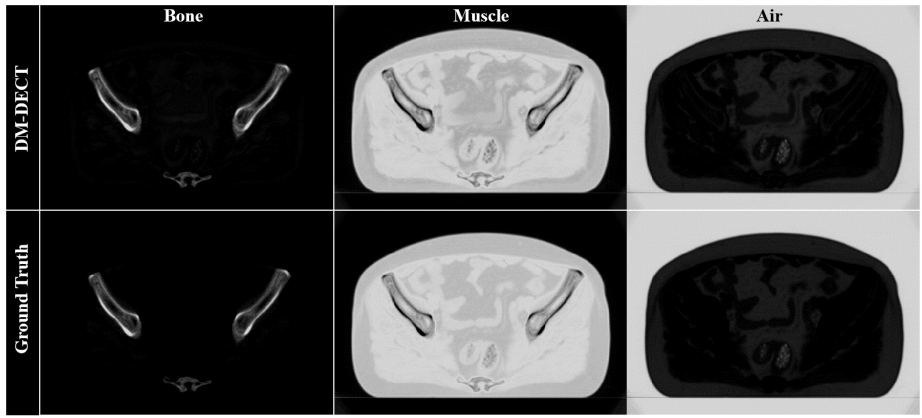

Figure 11. Decomposed volume fraction maps using DM-DECT and two-scan method as the ground truth. Display windows are [0 1.2] for all.

Line profiles along the red arrows are plotted on the third column of Fig. 10, showing great coherence in the reconstructed LAC values with the ground truth. The decomposed volume fraction maps are displayed in Fig. 11, where no visible difference appears between the DM-DECT results and the ground truth. SSIM value is 0.9241 between the DM-DECT decomposed images between the ground truth, showing the feasibility of the DM-DECT method in large-FOV imaging tasks.

## 4. Discussion and Conclusions

We propose a single-scan DECT method using the static detector modulation scheme. The beam modulator is installed in front of the detector to split the incident x-ray beams into high- and low-energy spectra. A pix2pix GAN model is introduced to achieve data restoration in the filtered sinogram domain. DECT images are obtained by the back-projection operation and multi-material decomposition is implemented in the sonogram domain using a FISTA-based algorithm.

The innovation of the proposed method compared with previously reported beam-modulated single-scan DECT methods (Lee *et al.*, 2017; Cho *et al.*, 2020; Petrongolo and Zhu, 2018) is mainly three-fold: small primary loss, deep learning-based data restoration scheme and easy manipulation. First of all, the proposed method employs the detector modulation scheme to reduce the dimension of the penumbra area compared with the source-side modulation methods by up to 80%. The significantly reserved data can significantly promote the accuracy of DECBCT reconstruction. Secondly, the data restoration model is introduced in this work to generate complete dual-energy data, enabling DM-DECBCT method to handle the clinically used half-fan and short-scan protocols which acquire only about 55% data of full scan. Finally, the proposed static modulator can be readily fabricated and attached to the detector surface without modification to the existing clinical CBCT hardware.

In the proposed DM-DECT method, projection data is sampled along the lateral direction. Subtle inconsistency along the lateral direction in the sinogram domain will induce severe ring artifacts in the CT images because of the derivative property of ramp filtration during the reconstruction process (Noo *et al.*, 2004; Zeng, 2007). To avoid the derivation on the generated data, data restoration is implemented in the filtered sinogram instead of directly in the sinogram domain (Lee *et al.*, 2021; Zhou *et al.*, 2021; Cao *et al.*, 2021). Material decomposition is performed on the forward projected DECT images, i.e., the simulated dual-energy sonogram data, due to the irreversibility property of the filtered sinogram for two reasons. Firstly, the ramp filtration can be factorized into the derivation and Hilbert transform in the spatial domain where the derivation is irreversible (Zeng, 2007; Noo *et al.*, 2004). Secondly, the imaginary part of the Fourier coefficient is discarded in the implementation of frequency-domain ramp filtration, which induces an irreversible operation.

Uniformly distributed metal strips with equal spacing and width are employed to construct the beam modulator for two reasons. Firstly, balanced high- and low-energy data acquisition is employed for their symmetrical status in the data restoration and image reconstruction. Secondly, equi-spaced strips guarantee high-frequency modulation across the whole object.

The grid width is empirically selected as 5 mm in this work. The strip width inside a specific interval should balance the tradeoff between higher modulation frequency and increased penumbra areas. Smaller width causes large amount of data loss in the penumbra areas. For example, 1 mm width causes up to 77% of data loss while 5-mm giving only 15% loss. Larger width, e.g., 10-, 15-, and 20-mm grids, causes less data loss but at the expense of degraded sampling performance of anatomical structures.

There are still some aspects worth further study to boost the clinical practice of the proposed method. Firstly, due to the data consistency requirement of deep learning methods, individual patient data acquisition is required for model training of different DM-DECT systems in clinical practice (Zhou *et al.*, 2021). In current work, we expect to collect at least 20 number of patient full data set to finalize the model training. To reduce the required data size for the model training, we will incorporate the transfer learning technique (Pan and Yang, 2009) due to its pretrain and retrain strategy for data set reduction. A system-independent model can be pretrained using opened clinical CT data to learn the anatomical structural information. The pretrained model can learn the spectral information for a specific system by being retrained with only 10 patient data acquired on the specific machine, which means reduced total dose exposure during the training data acquisition. Secondly, anatomical and spectral information in the sinogram and image domains is disregarded in the proposed single-domain learning method because only filtered sinogram-domain loss is utilized. To fully exploit all the information in each domain, multi-domain learning scheme can be developed (Peng *et al.*, 2020). We plan to introduce two domain-transform networks to the data restoration model. One domain-transform module performs ramp filtration to transform the sinogram to filtered sinogram. The other module performs back-projection operation to generate the CT images. Adversarial and $L_1$ losses in all the three domains are utilized to further improve the performance of data restoration. Thirdly, scatter contamination is non-neglectable in the cone-beam scan, leading to altered pixel values in the acquired data and biased output. A possible way to overcome this issue is to perform scatter correction in original projection data before feeding the modulated data into the deep learning model. Several comprehensive investigations have been performed to scatter correction in prior publications (Jiang *et al.*, 2019; Niu and Zhu, 2011). A practical scatter correction scheme in DECBCT image will incorporate the modulation effect to scatter photons. Finally, beam hardening artifacts cannot be eliminated in this work since the proposed material decomposition is performed in sinogram instead of projection domain. Polychromatic spectral information is thus not included in current implementation. To address this issue, we will include the spectral information in the imaging model and implement the material decomposition in the projection domain (Long and Fessler, 2014).